\documentstyle[graphics,graphicx,times,color]{mn}
\title[New protostars in L\,1630]{Dense cores in the
L1630 molecular cloud: discovering new protostars with SCUBA}

\author[Robin R. Phillips, Andy G. Gibb and Leslie T. Little]
       {Robin R. Phillips$^1$\thanks{Present address: Joint Astronomy
       Centre, 660 N. A`oh\=ok\=u Place, Hilo, HI 96720, USA}, Andy
       G. Gibb$^2$\thanks{Present
       address: Department of Astronomy, University of Maryland,
       College Park, MD 20742, USA} and Leslie T. Little$^1$ \\
        $^1$ Electronic Engineering Laboratory, University of Kent at
       Canterbury, Canterbury, CT2 7NT \\
	$^2$ Department of Physics and Astronomy, University of Leeds,
       Leeds, LS2 9JT}

\date{Accepted 2000. 
      Received 1999}

\pagerange{\pageref{firstpage}--\pageref{lastpage}}
\pubyear{2000}

\begin{document}

\maketitle
\label{firstpage}

\begin{abstract} 

\noindent
Maps of the 450 $\mu$m and 850 $\mu$m dust continuum emission from
three star-forming condensations within the Lynds 1630 molecular
cloud, made with the SCUBA bolometer array, reveal the presence of
four new submillimetre sources, each of a few solar masses, two of
which are probably Class I, two Class 0, as well as several sources
whose existence was previously known. The sources are located in
filaments and appear elongated when observed at 450 $\mu$m. They
likely have dust temperatures in the range 10 to
20 K, in good agreement with previous ammonia temperature
estimates. Attempts to fit their structures with power-law and
Gaussian density distributions suggest that the central distribution
is flatter than expected for a simple singular isothermal sphere.

Although the statistics are poor, our results suggest that the ratio
of `protostellar core' mass to total virial mass may be similar for
both large and small condensations.
\end{abstract}


\section{Introduction}

The Orion molecular clouds are the subject of extensive investigation
on account of their proximity (400--500 pc) and the formation within
them of stars of a variety of masses. An important unbiased survey of
molecular condensations within the clouds is that of Lada, Bally and
Stark (1991; hereafter LBS), who mapped the Orion B (L\,1630)
cloud with an angular resolution of 1.7 arcmin, observing emission
from the $J$=2--1 transition of interstellar CS. Excitation of this 
transition requires  molecular hydrogen densities of order 10$^4$--10$^5$
cm$^{-3}$. Several of these condensations have themselves been mapped
with substantially higher resolution and in molecular transitions
requiring still higher densities for excitation (Zhou et al. 1991;
Lis, Carlstrom \& Phillips 1991; Chandler, Moore \& Emerson 1992).

Gibb \& Heaton (1993; hereafter GH93) and Gibb~et~al. (1995; Paper I)
mapped 6 of the LBS condensations in the $J$=3--2 or $J$=4--3
transitions of interstellar HCO$^+$. These revealed an apparently
filamentary structure containing embedded cores whose virial masses
were typically a few solar masses.

Paper I mapped the dust continuum emission from several of the cores
in LBS23, while Gibb \& Little (1998; hereafter GL98) mapped the whole
of LBS23 in $J$=2--1 C$^{18}$O. Dust-derived masses for the cores were
in good agreement with their virial masses derived from HCO$^+$. The
emission peaks evident in HCO$^+$ usually correlated well with those
of the dust continuum but were faint or missing in C$^{18}$O. GL98
considered that these results implied depletion of the C$^{18}$O by
factors of between 10--50. The lack of correlation between dust
continuum and C$^{18}$O was similar to that found in the hotter source
NGC2024 \cite{mwm}.  The simple interpretation as depletion in NGC2024
\cite{msh} has been questioned by
Chandler~\&~Carlstrom~\shortcite{cjc} on the grounds that (a) the dust
opacity law in the cores may be non-standard, (b) dust emission may
become optically thick at the short submm wavelengths, but erroneously
interpreted as thin, (c) the cores may contain unresolved components
which are highly optically thick in C$^{18}$O but interpreted as thin
so the true molecular abundance is much higher.
 
The recent availability of the Submillimetre Common User Bolometer
Array (SCUBA: Holland~et~al. 1999) permits rapid mapping of dust
emission from the LBS condensations with greatly improved sensitivity.
This means that large areas within them can be mapped to determine the
gross dust distribution and further protostellar cores may be readily
identified. We here describe the results of a pilot study in which
selected areas of our LBS condensations have been mapped with SCUBA.

\begin{figure*}
\resizebox{\hsize}{!}{
\includegraphics{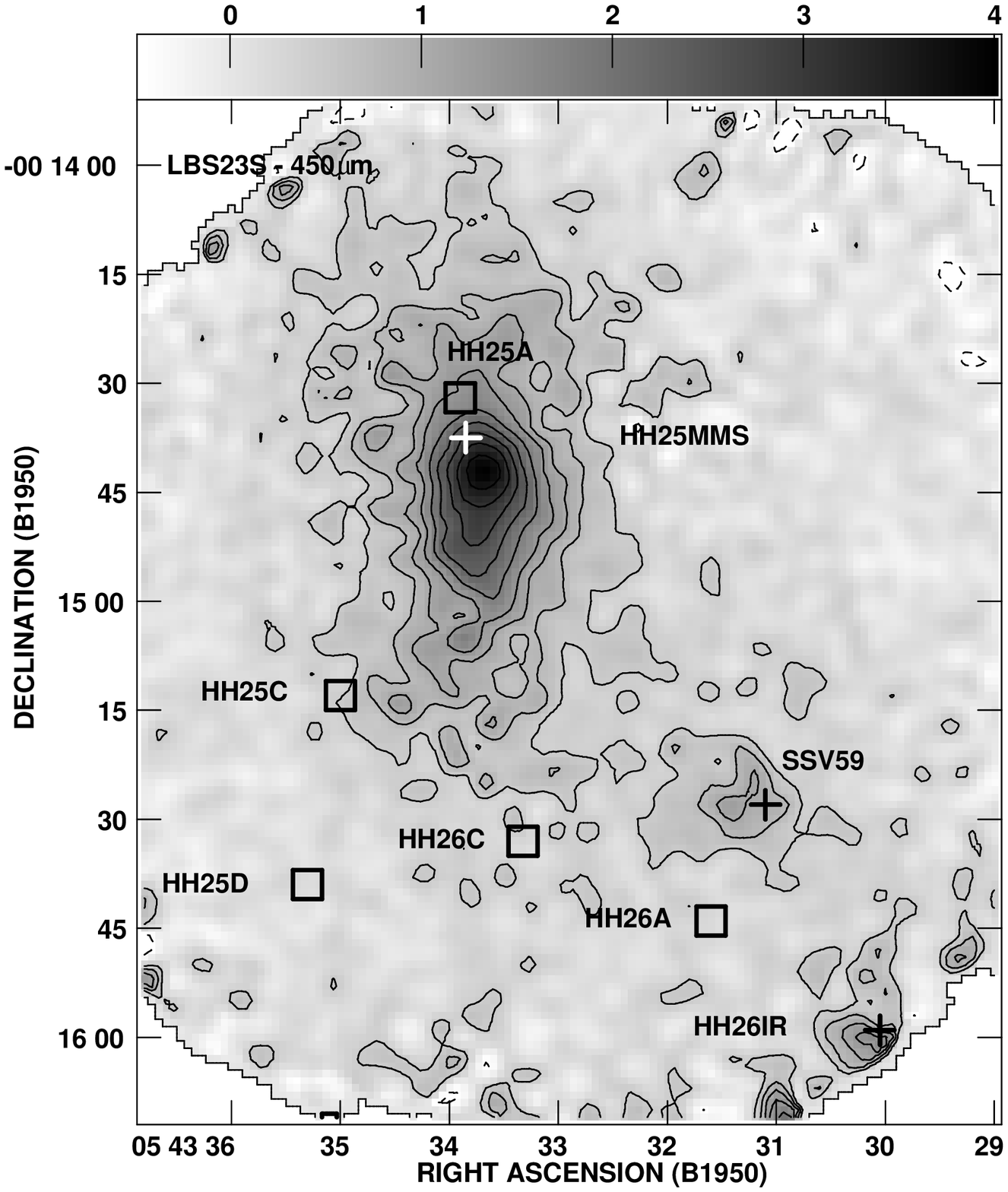}
\includegraphics{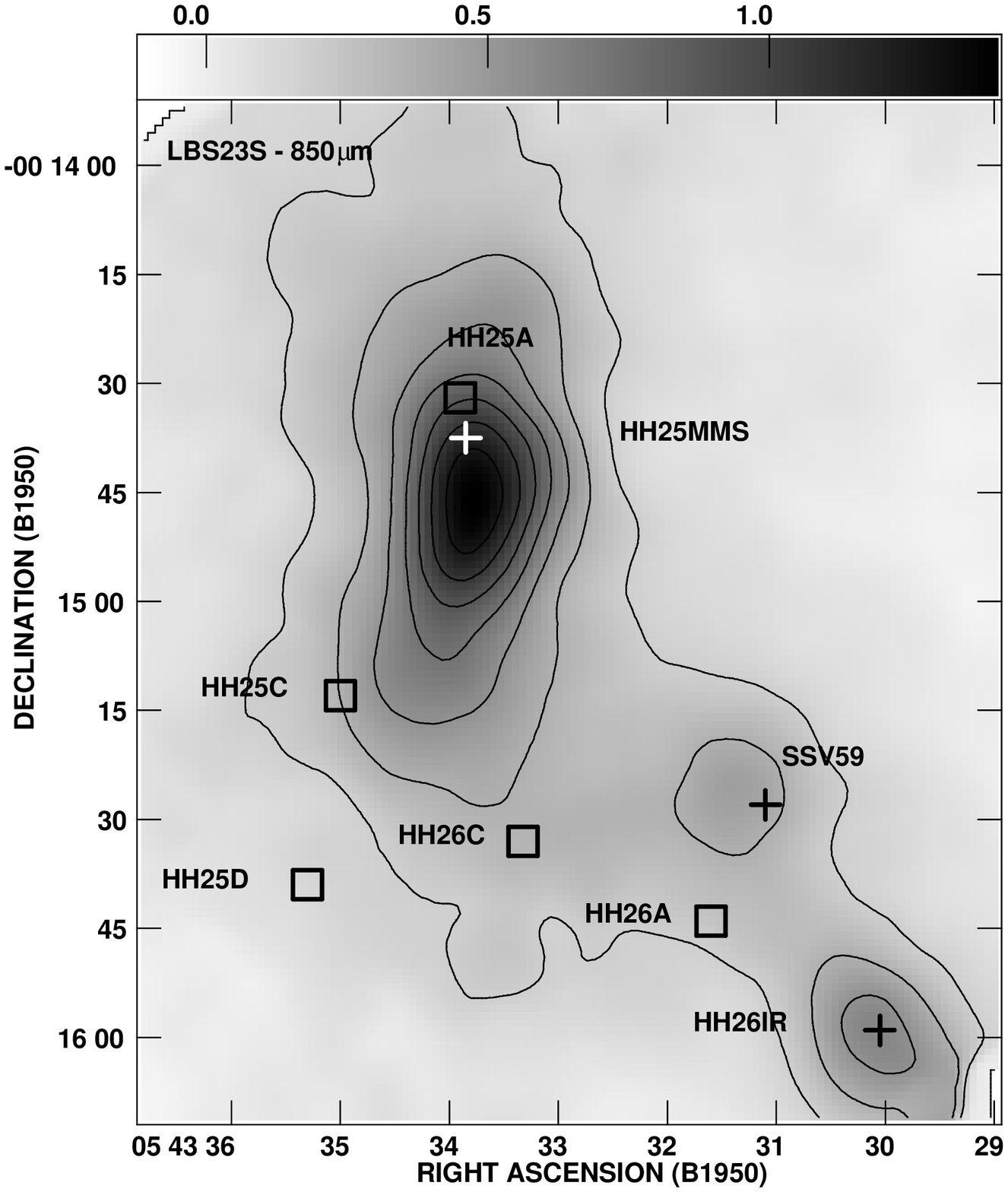}
}
\caption[]{SCUBA maps of LBS23S at 450 $\mu$m (left) and 850 $\mu$m
(right). The known sources HH25MMS, SSV59 and HH26IR are marked by
crosses, while the components of HH25 and HH26 from Davis et
al. (1997) are marked by open squares. The greyscale is in Jy/beam and
is shown as a bar across the top of each figure, extending from
$-3\sigma$ to the maximum (4.0 Jy/beam at 450 $\mu$m and 1.4 Jy/beam
at 850 $\mu$m). Contours are --3, 3, 5, 7, 9, 12, 15, 18, 21, 24 (450
$\mu$m) and 5, 10, 15, 20, 25, 30, 35, 40 (850 $\mu$m) times the noise
level of 150 mJy per beam and 36 mJy per beam respectively.
\label{fig:lbs23s}}
\end{figure*}

The motivation for this study was:
\begin{description}
\item[(a)] to search for dust components near, for example, outflow 
sources, near-infrared sources, and Herbig-Haro objects, where  
they have not previously been detected, and for new protostellar objects.
\item[(b)] to compare the observed dust emission with models of the
structure of protostellar cores such as those of Shu~\shortcite{shu},
or Crutcher~et~al.~\shortcite{cmt}.  The simultaneous mapping of the
cores at 850 and 450 $\mu$ms facilitates some separation of dust mass
and temperature.
\item[(c)] to see whether the morphology of large scale filaments
fragmenting into protostellar cores evident in e.g. NGC2024 and LBS23 extends
to smaller condensations.
\item[(d)] to compare the numbers of protostellar cores in large and 
small condensations in the same molecular cloud and to see if stars form 
preferentially in the larger condensations.
\item[(e)] to map fields for which detailed HCO$^+$ and C$^{18}$O maps exist 
already, to determine molecular depletion in different source components. A
detailed analysis will follow in a later paper. 
\end{description}

\section{Observations}

%\begin{table}
%\begin{center}
%\caption[]{Summary of SCUBA observations of LBS cores. The virial mass
%for each is taken from LBS. Noise levels are given per beam. The dust
%masses are calculated from the 3$\sigma$ noise level in the 850-$\mu$m
%images. \label{tab:scubaobs}}
%\begin{tabular}{lccccc}%c}
%Source  & Virial & 850 $\mu$m &
%        \multicolumn{2}{c}{Mass(M$_\odot$)/beam} & 450 $\mu$m \\%& Protostellar \\
%        &  Mass  & noise      &  \multicolumn{2}{c}{limits (3$\sigma$)}       &    noise   \\%&   clumps    \\
%        & (M$_\odot$)         &      &  10 K   &  20 K    &    \\%& \\
%%        &        &            &       &    -     & \\ 
%\hline
%LBS7      & 16     &   80 & 1.5 &  0.42  &  -- \\%& No \\
%LBS13     &  55    &   83 & 1.5 &  0.45  &  -- \\%& No \\
%LBS16     &  36    &   80 & 1.5 &  0.42  &  -- \\%& No \\
%LBS17     & 265    &      &     &        &     \\%&    \\
%~~~~~~~H  &        &   35 & 0.6 &  0.19  & 130 \\%& Yes\\
%LBS18     &  15    &   33 & 0.6 &  0.18  & 150 \\%& Yes\\
%LBS23     & 230    &      &     &        &     \\%&    \\
%~~~~~~~N  &        &   30 & 0.6 &  0.15  & 160 \\%& Yes\\
%~~~~~~~S  &        &   36 & 0.6 &  0.19  & 150 \\%& Yes\\ 
%\end{tabular}
%\end{center}
%\end{table}     

\begin{figure*}
\begin{center}
\includegraphics[scale=0.55]{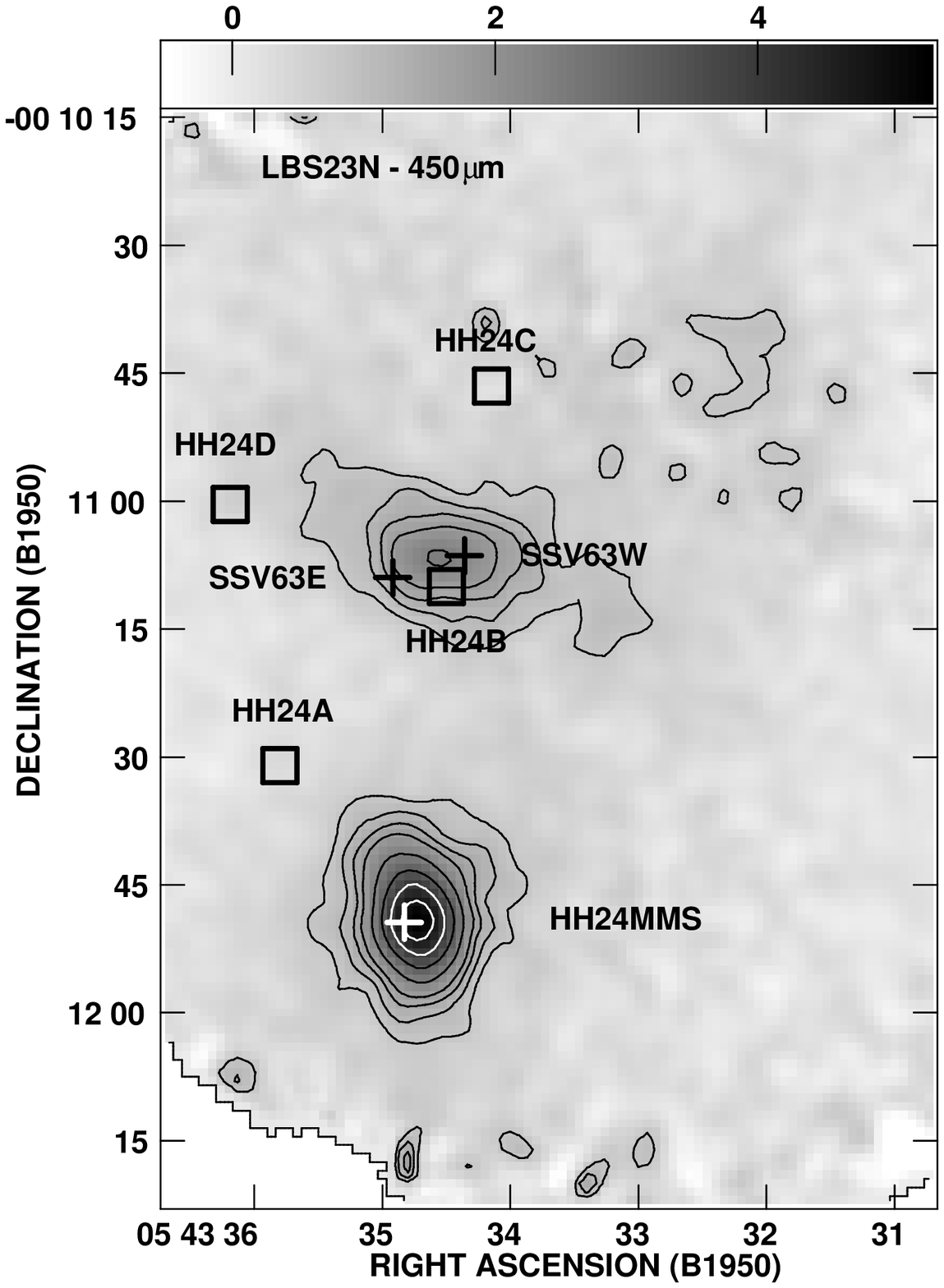}
\includegraphics[scale=0.55]{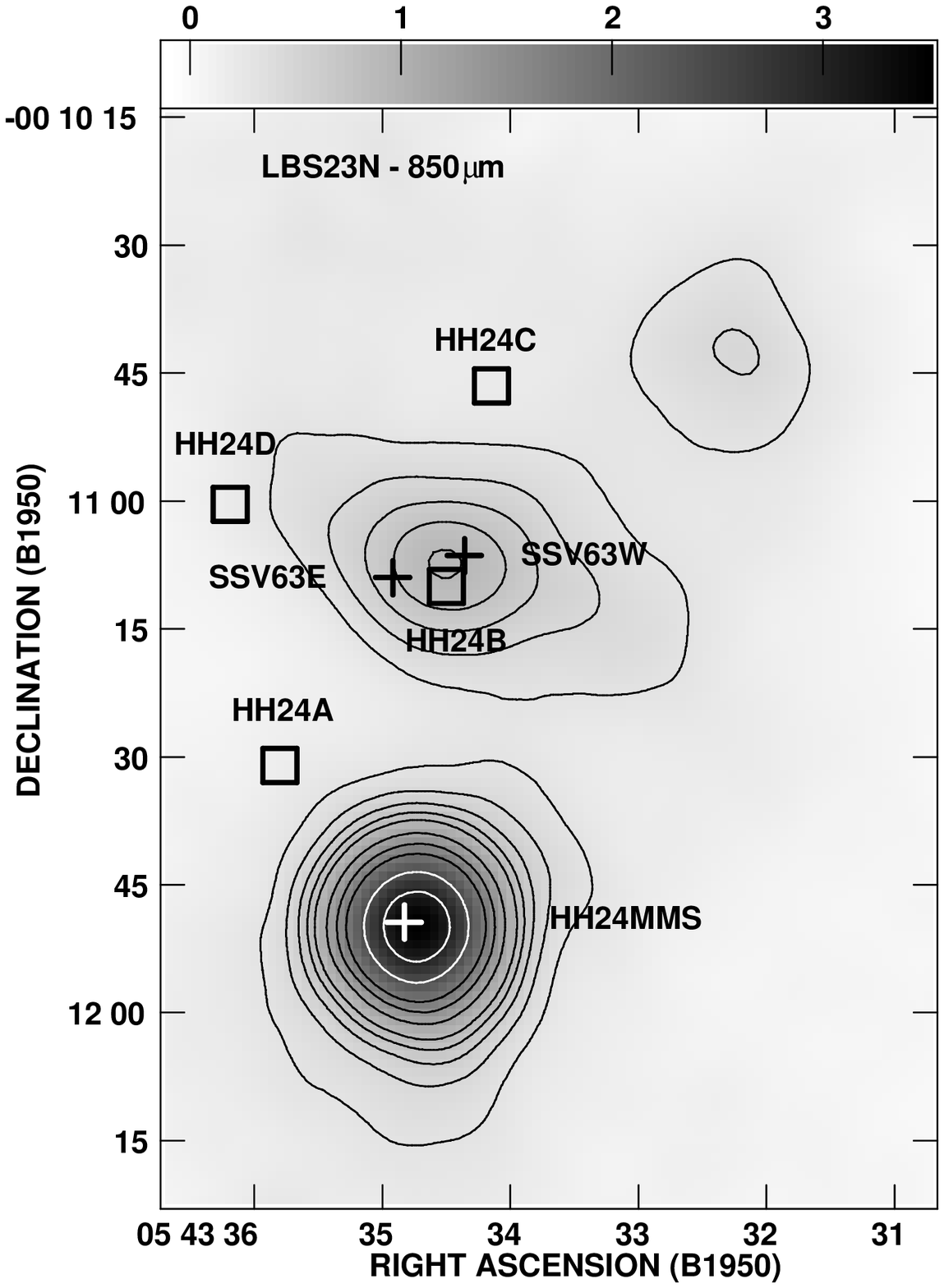}
\caption[]{SCUBA maps of LBS23N at 450 $\mu$m (left) and 850 $\mu$m
(right). The known sources HH24MMS, SSV63E and SSV63W are marked by
crosses while the Herbig-Haro objects HH24A--D are marked by open
squares. The greyscale is in Jy/beam and is shown as a bar across the
top of each figure, extending from $-3\sigma$ to the maximum (5.3
Jy/beam at 450 $\mu$m and 2.5 Jy/beam at 850 $\mu$m).  Contours are
--3, 3, 5, 7, 9, 11, 15, 20, 25, 30 (450 $\mu$m) and 10, 15, 20, 25,
30, 40, 50, 60, 80, 100 (850 $\mu$m) times the noise level of 160 mJy
per beam and 30 mJy per beam respectively. \label{fig:lbs23n}}
\resizebox{\hsize}{!}{
\includegraphics[scale=0.55]{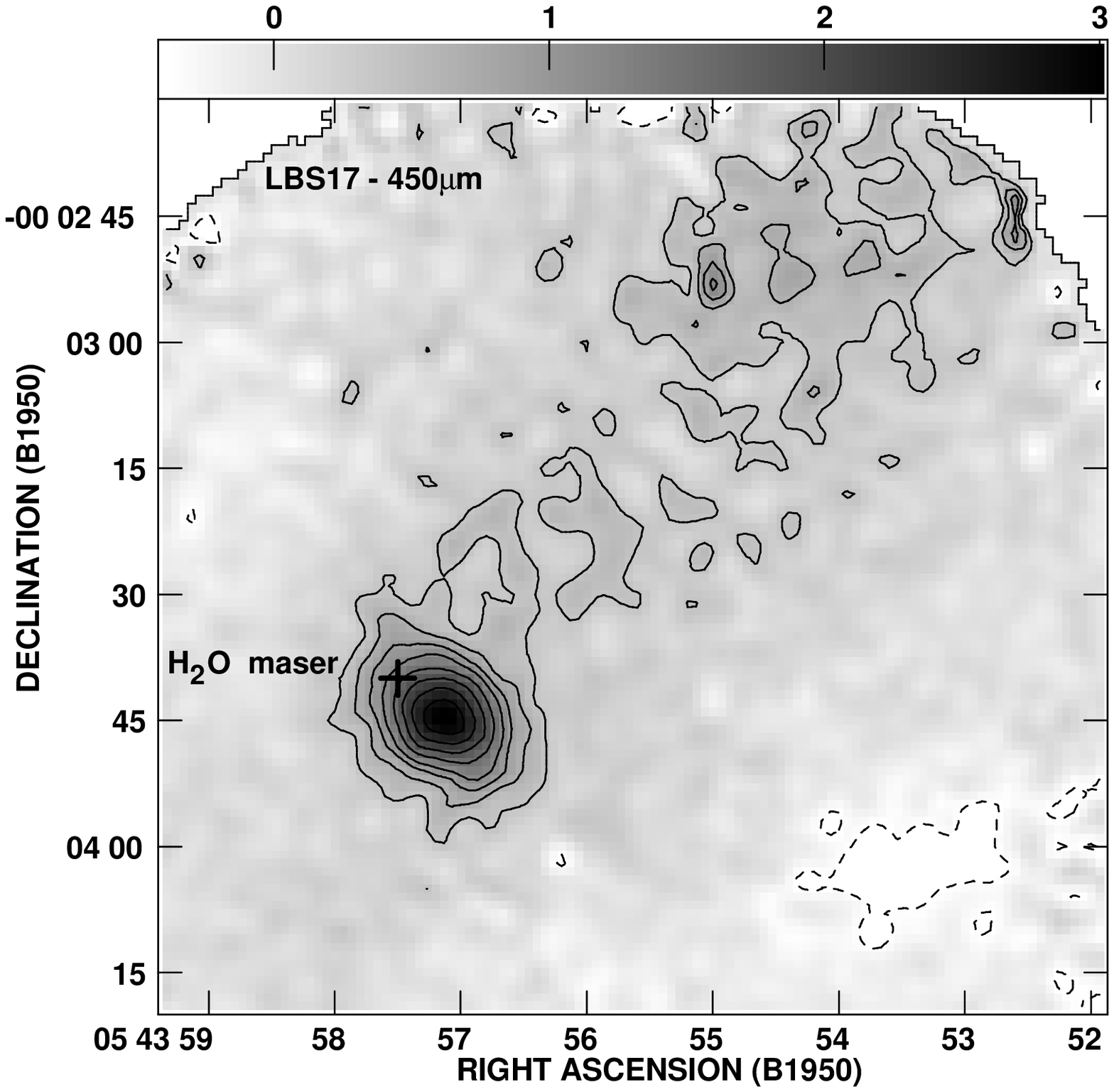}
\includegraphics[scale=0.55]{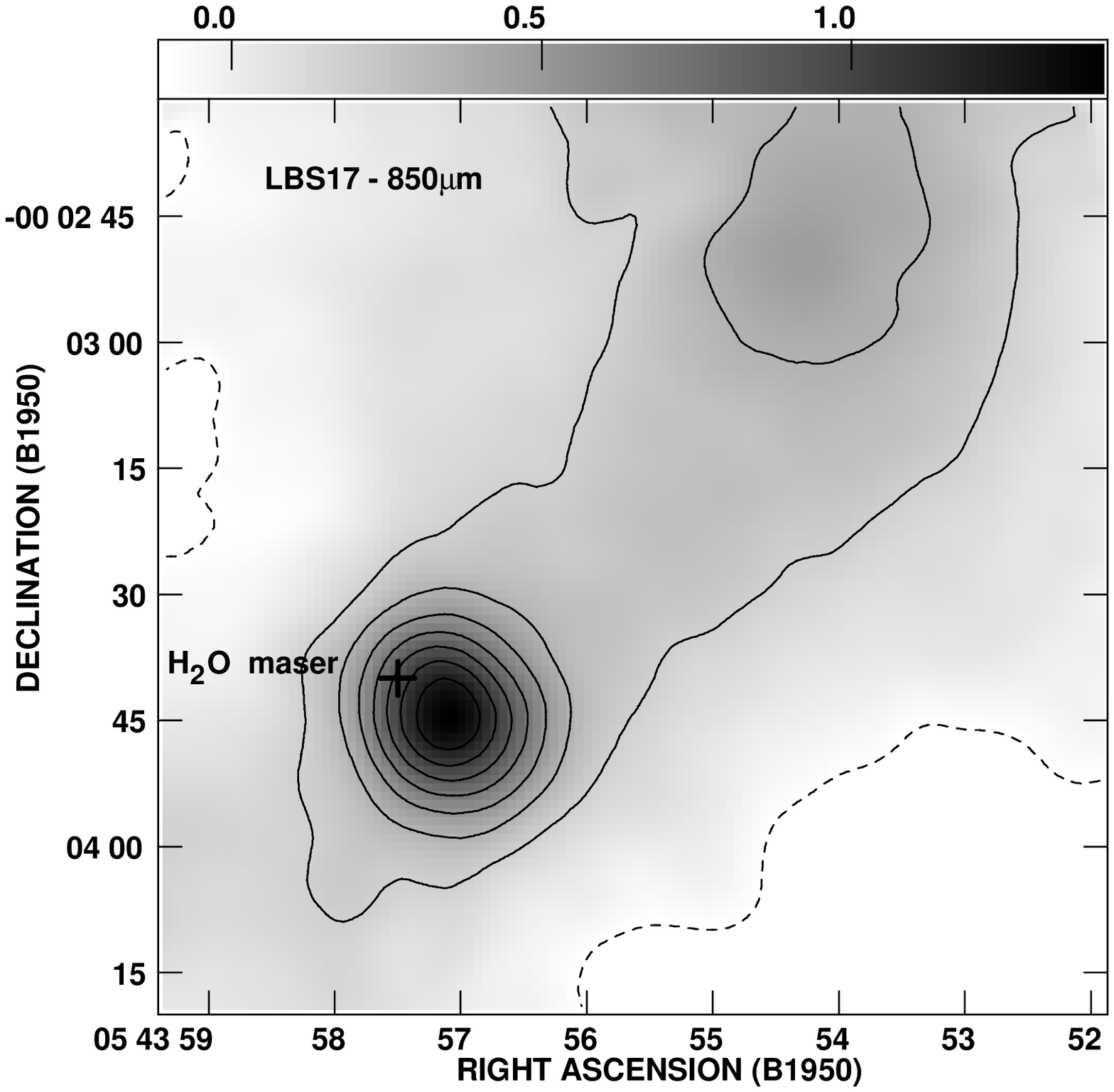}
}
\caption[]{SCUBA maps of LBS17 at 450 $\mu$m (left) and 850 $\mu$m
(right). The water maser of Haschick et al. (1983) is marked by a
cross. The greyscale is in Jy/beam and is shown as a bar across the
top of each figure, extending from $-3\sigma$ to the maximum (3.0
Jy/beam at 450 $\mu$m and 1.4 Jy/beam at 850 $\mu$m). Contours are
--3, 3, 5, 7, 9, 12, 15, 18, 21, 24 (450 $\mu$m) and --3, 5, 10, 15,
20, 25, 30, 35, 40 (850 $\mu$m) times the noise level of 130 mJy per
beam and 35 mJy per beam respectively. \label{fig:lbs17}}
\end{center}
\end{figure*}

\begin{figure*}
\resizebox{\hsize}{!}{
\includegraphics[scale=0.55]{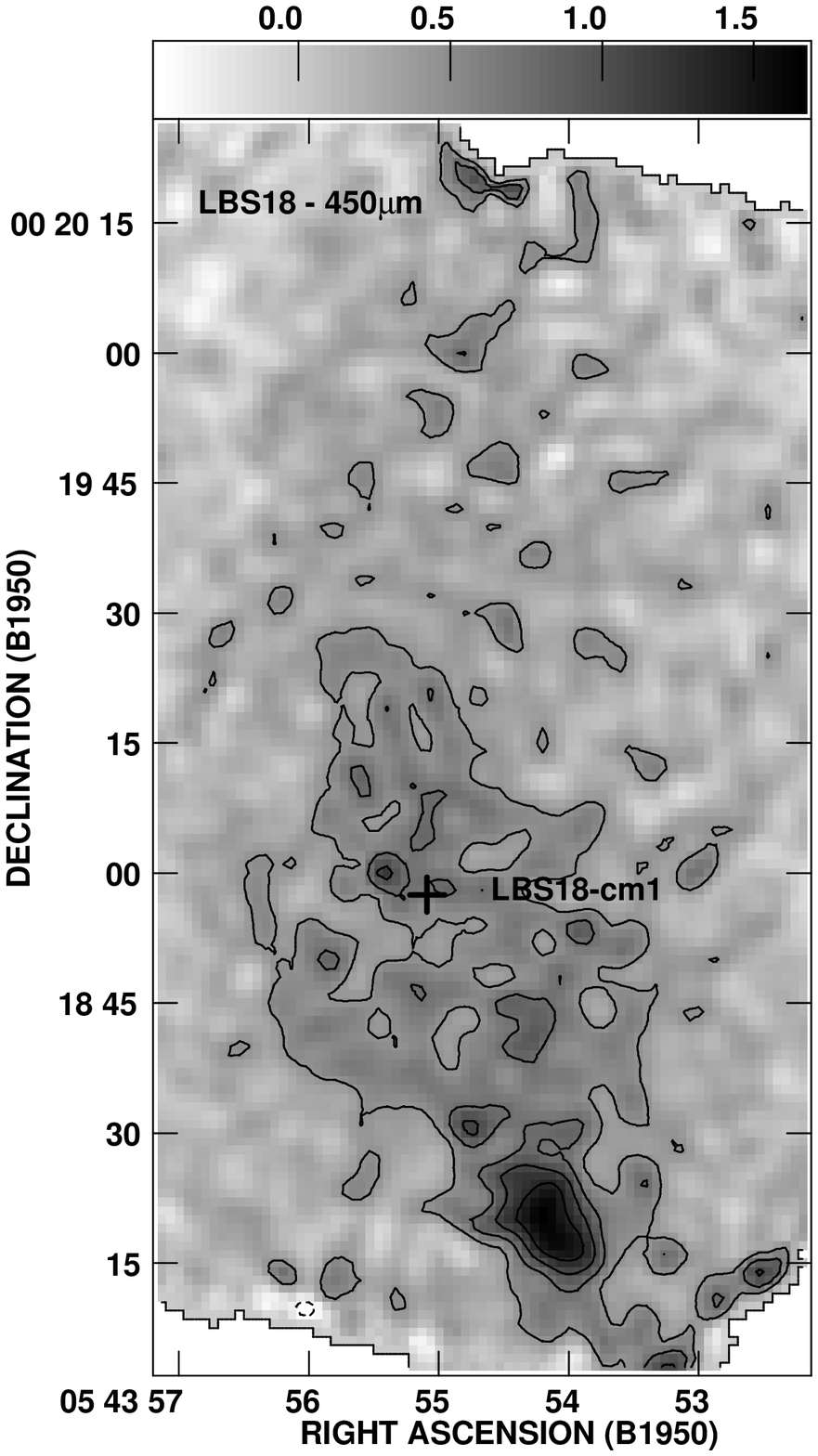}
\includegraphics[scale=0.55]{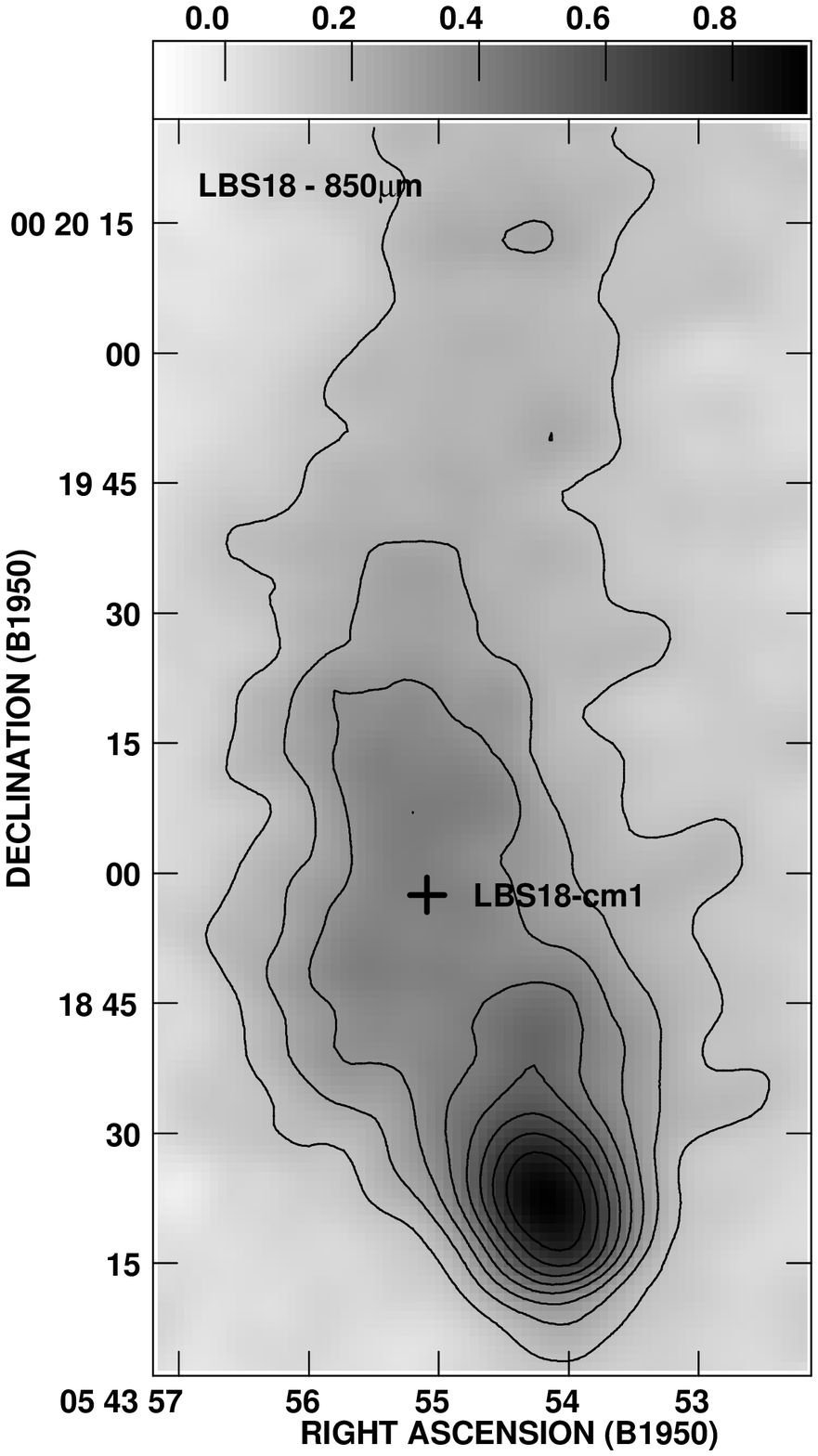}
}
\caption[]{SCUBA maps of LBS18 at 450 $\mu$m (left) and 850 $\mu$m
(right). The 3.5-cm source detected by Gibb (1999) is marked by a
cross. The greyscale is in Jy/beam and is shown as a bar across the
top of each figure, extending from $-3\sigma$ to the maximum (1.7
Jy/beam at 450 $\mu$m and 0.91 Jy/beam at 850 $\mu$m). Contours are
--3, 3, 5, 7, 9, 12, 15, 18, 21 (450 $\mu$m) and 5, 8, 11, 14, 17, 20,
23, 26 (850 $\mu$m) times the noise level of 150 mJy per beam and 33
mJy per beam respectively. \label{fig:lbs18}}
\end{figure*}

The observations were made using the submillimetre bolometer array
receiver SCUBA \cite{crg} operated in service mode at the James Clerk
Maxwell Telescope over the period 24 to 26 August 1997 and
20 and 23 September 1997. The 32 element 850 $\mu$m and 96 element 450
$\mu$m arrays were used in `jiggle-mode' to make fully-sampled maps of
2.2-arcmin diameter-fields in 6 of the LBS cores. The chop throw was
120 arcsec in azimuth. The flux scales were established by
observations of Uranus at the start of each observing session and
skydips were used to calculate the zenith optical depth at the start
and end of each session.

The night of 26 August was very humid and calibration produced results
that differed by of order 50 per cent (lower) compared with the
previous and following nights. The weather on the nights of 20 and 23 September
 was poor and the 450 $\mu$m data could not be used. Thus the
images presented were constructed from data recorded on the nights of
24 and 25 August 1997 only. On these two nights, the estimated
calibration uncertainty is of order 10 per cent at 850 $\mu$m and 30
per cent at 450 $\mu$m. The beam of the telescope was determined to
have full-width at half-maximum (FWHM) diameter of 15.0$\pm$0.5 arcsec
at 850 $\mu$m and 8.1$\pm$0.2 arcsec at 450 $\mu$m. The total time on
each source was of order one hour, except for LBS7, 13 and 16. Due to
poor weather the time on source for these fields was only 21 minutes.

The typical noise level at 850 $\mu$m and 450 $\mu$m was 34 mJy per
beam and 150 mJy per beam respectively for LBS17, 18 and the two
fields in LBS23. This translates to a 3$\sigma$ mass sensitivity of
0.6 M$_\odot$ for temperatures of 10 K, assuming that the dust
emissivity is as described by Hildebrand~\shortcite{hildebrand}. The
noise level for LBS7, 13 and 16 was higher at $\sim$80 mJy per beam at
850 $\mu$m; no emission was detected in these fields.
 
Table~\ref{tab:class} summarises the fluxes and positions of the dust
emission sources detected in the LBS cores.

\section{Results}

Figs. 1 to 4 show the SCUBA maps at 450 and 850 $\mu$m for the four
LBS fields in which emission was detected.  Table~\ref{tab:class} lists the
parameters derived from two-dimensional gaussian fits (using the task
{\sc imfit} in {\sc aips}) for the sources detected in the LBS17, 18
and 23 fields. The source dimensions have not been deconvolved.  The
(formal) uncertainties in the positions and dimensions arising from
the fitting procedure are of order 0.5 arcsec. The main features of
each source will now be discussed in turn.

\subsection{LBS23S}

This field is centred on the class 0 source HH25MMS \cite{gd98} and
contains the class I infrared sources SSV59 and HH26IR \cite{d97}. The
purpose of mapping this region was to compare the SCUBA maps of
HH25MMS (Fig. 1) with the existing UKT14 map (Paper I) and also to
search for the driving source of the southern bipolar outflow in
HH24-26 (GH93). The morphology of HH25MMS is consistent in both maps,
and the maximum lies within a couple of arcsec of the 3.4-mm peak of
Choi et al.~\shortcite{choi}. Two further weaker components are
evident in Fig. 1; they are associated with HH26IR, which is the
source of the southern outflow (GH93), and SSV59, which also drives a
molecular outflow \cite{gd97}. Edge effects make it difficult to make
accurate measurements of the flux density of HH26IR as we have not
observed the entire source. Recent 1300-$\mu$m and 350-$\mu$m
observations of the whole of LBS23 by Lis, Menten \& Zylka (1999)
reveal a structure consistent with what we observe in Figs. 1 and
2. They show that HH26IR also appears to be embedded within a clump
elongated perpendicular to the CO flow.

A curious point is that the radio continuum source and origin for the
HH25MMS molecular outflow is not coincident with the submillimetre
continuum peak, being offset by $\sim$6 arcsec in declination,
somewhat larger than the mean pointing error (3 arcsec). If genuine,
this observation may indicate the presence of a second submillimetre
source within LBS23S, which may be younger than HH25MMS (on account of
there being no observable outflow). Alternatively, it may be a
hot-spot heated by the impact of the outflow from HH25MMS. Clearly
higher resolution observations are needed to clarify this issue.  Lis
et al. (1999) observe this in their 1300-$\mu$m map, but it seems less
pronounced in their 350-$\mu$m data.

\begin{table*}
\begin{center}
\caption[]{Source parameters, including identification where
known. The source dimensions were derived from 2-dimensional gaussian
fits and list the FWHM of the major and minor axes (in arcsec) and the
position angle of the major axis in degrees east of north. Angles are
rounded to the nearest integer. The last three columns list the flux
density (in Jy) within an aperture of either 8 or 15 arcsec.
\label{tab:class}}
\begin{tabular}{llcccccccc} 
LBS & Source & Source Position & $a_{850} \times b_{850}$ & PA &
$a_{450} \times b_{450}$ 
& PA & $F(8'')$ & \multicolumn{2}{c}{$F(15'')$}  \\
core & & RA(1950.0), Dec(1950.0) & arcsec$^2$ & degrees
& arcsec$^2$ & degrees & 450 $\mu$m & 850 $\mu$m & 450 $\mu$m \\
 \hline

17 & LBS17H  & 05$^{\rm h}$43$^{\rm m}$57.1$^{\rm s}$, 
--00$^\circ$03$'$44.3$''$ & 18$\times$16 & 26$\pm$2 & 12$\times$9 & 49$\pm$3 &
    3.0            &     1.4         &  5.4 \\

17 & LBS17F  & 05$^{\rm h}$43$^{\rm m}$54.2$^{\rm s}$, 
--00$^\circ$02$'$48.2$''$ & 32$\times$23 & --27$\pm$3 & -- & -- &
    --            &     0.4         & -- \\

18 & LBS18S  & 05$^{\rm h}$43$^{\rm m}$54.2$^{\rm s}$, 
+00$^\circ$18$'$23.1$''$  & 26$\times$20 & 22$\pm$4 & 14$\times$10 & 31$\pm$3 &
    1.7            &     0.9         &  3.5   \\

23 & HH25MMS & 05$^{\rm h}$43$^{\rm m}$33.8$^{\rm s}$,
--00$^\circ$14$'$46.5$''$ & 29$\times$15 & --8$\pm$1 & 15$\times$13 &
--1$\pm$9 &
    3.7            &     1.4         &  8.7  \\

23 & SSV59   & 05$^{\rm h}$43$^{\rm m}$31.4$^{\rm s}$,
--00$^\circ$15$'$27.3$''$ & 25$\times$22 & 13$\pm$6 & 11$\times$10 &
79$\pm$43 &
    1.1            &     0.4         &  2.2   \\

23 & HH26IR  & 05$^{\rm h}$43$^{\rm m}$30.1$^{\rm s}$,
--00$^\circ$16$'$00.1$''$ & 27$\times$17 & 42$\pm$2 & 9$\times$8 &
--57$\pm$64 &
    1.4            &     0.6         &  --  \\

23 & HH24MMS & 05$^{\rm h}$43$^{\rm m}$34.7$^{\rm s}$,
--00$^\circ$11$'$48.8$''$ & 17$\times$16 & 2$\pm$5 & 13$\times$9 & 9$\pm$2 &
    5.1            &     3.5         &  9.6   \\

23 & SSV63   & 05$^{\rm h}$43$^{\rm m}$34.4$^{\rm s}$, 
--00$^\circ$11$'$07.3$''$ & 23$\times$17 & 63$\pm$5 & 16$\times$10 &
79$\pm$3 &
    2.4            &     1.0         &  5.1  \\ 

23 & HH24NW  & 05$^{\rm h}$43$^{\rm m}$32.3$^{\rm s}$, 
--00$^\circ$10$'$43.1$''$ & 24$\times$21 & 21$\pm$7 & -- & -- &
    --            &      0.5         &  --  \\ 
\end{tabular}
\end{center}
\end{table*}

\subsection{LBS23N}

This field is centred on the HH24 complex, and includes the class 0
protostar HH24MMS. The two components HH24MMS and SSV63W detected at
1300 $\mu$m by Chini~et~al.~\shortcite{ckh} are clearly mapped
(Fig. 2). There is good agreement between the 850 $\mu$m SCUBA and 800
$\mu$m UKT14 (Paper I) peak fluxes for the unresolved source
HH24MMS. The 450 $\mu$m map reveals the source to be extended in a
N--S direction. While the flux densities are also in good agreement
with the UKT14 data of Ward-Thompson et al.~\shortcite{wck} the SCUBA
450 $\mu$m observations show the source as much more compact. We note
that the direction of the compact jet discovered by Bontemps,
Ward-Thompson \& Andr\'e~\shortcite{bwa} is neither parallel nor
perpendicular to the elongation at 450 $\mu$m. A new weak source is
seen 40 arcsec north-west of SSV63 which shows up weakly at 450 $\mu$m
as well. We label this clump HH24NW (Table~\ref{tab:class}) although
its parameters are not particularly well determined. Lis et al. (1999)
also detect this source.

\subsection{LBS17}

Fig. 3 presents the submillimetre map of the south-eastern portion of
the HCO$^+$ ridge of Paper I. The $J$=3--2 HCO$^+$ map produced a peak
largely missing in C$^{18}$O (Paper I; Gibb \& Little 2000). However,
this peak stands out as a dust source in the SCUBA map. There is a
bipolar outflow associated with this source, observable in HCO$^+$ and
CO (Gibb \& Little 2000).  The elongation of the dust source,
observable at 450 $\mu$m, is perpendicular to the direction of the
bipolar outflow. No IRAS, optical or near-infrared emission appears to
be associated with the source -- so it is likely to be young, possibly
protostellar. Gibb \& Little (2000) showed that it satisfies the
criteria for being a class 0 candidate. We refer to this source as
LBS17H, to distinguish it from another source detected in the LBS17
condensation to the north of the area covered by our map by
Launhardt~et~al.~\shortcite{lmh}. This second source lies near a peak
of our HCO$^+$ map (Paper I). A ridge of dust emission (Fig. 3) is
perceived to run NW towards an extended plateau leading on to this
peak, a feature also evident in HCO$^+$.  A poorly-defined peak is
seen in the northwest of the 850-$\mu$m image although this may be an
artefact of the mapping procedure. However, it is coincident with one
of the HCO$^+$ clumps of Paper I, clump B6 although in keeping with
the revised nomenclature we label it LBS17F in
Table~\ref{tab:class}. A more extensive map is required to determine
its true parameters.

\subsection{LBS18}

LBS18 has one of the smallest virial masses of the LBS cores (14
M$_\odot$). In dust emission (Fig. 4), as in $J$=3--2 HCO$^+$ (Paper
I) and $J$=2--1 C$^{18}$O (Little et al., in prep), the source has a
`cometary' appearance, showing a N--S elongation with a compact
component at one end, which could be
protostellar. Gibb~\shortcite{g99} detected a radio continuum source
within this core using the VLA but it lies within the main north-south
ridge, and is not coincident with the compact component.

\begin{table} 
\begin{center}
\caption[]{Masses derived from simple isothermal analysis assuming a
temperature estimated from previous NH$_3$ data and an emissivity
index of 2. The mass within 8 arcsec was calculated from the 450
$\mu$m flux density; that within 15 arcsec from the 850 $\mu$m flux
density. The Class label is that defined by Lada~\shortcite{l91} and
updated by Andr\'e, Ward-Thompson \& Barsony~\shortcite{awb93}.
\label{tab:simpleanalysis}}
\begin{tabular}{lccccc}
Source   & $T_{\rm d}$   &   $M$ (M$_\odot$) & $M$ (M$_\odot$) &
         Class  &  Outflow     \\
         &   (K)   & (8 arcsec)& (15 arcsec)     &        &     \\ 
\hline
%LBS17H  & 9$\pm$2 & 5.8$^{-2.9}_{+10.4}$ & 10.6$^{-3.6}_{+8.9}$ & 0 & Y \\ 
%LBS18S  & 9$\pm$2 & 3.3$^{-1.6}_{+5.9}$ &  6.8$^{-2.3}_{+5.7}$ & ? & N? \\   
%HH25MMS & 14$^{+6}_{-4}$ & 1.8$^{-1.0}_{+3.1}$ & 4.5$^{-2.0}_{+4.0}$ & 0 & Y \\
%SSV59   & 11$^{+4}_{-2}$ & 1.1$^{-0.6}_{+1.0}$ & 2.0$^{-0.9}_{+1.0}$ & I & Y \\
%%HH26IR  & 15$\pm$5 &    --     &                 &  I     &  Y  \\
%HH24MMS & 8$\pm$1  & 16$^{-6}_{+12}$ & 35$^{-8}_{+14}$ & 0 & Y \\
%SSV63   & 11$^{+4}_{-2}$ & 2.4$^{-1.4}_{+2.2}$ & 5.0$^{-2.2}_{+2.6}$ & I & Y \\
LBS17H  & 15 & 0.9 & 2.1 & 0 & Y \\ 
LBS18S  & 15 & 0.5 & 1.4 & ? & N? \\   
HH25MMS & 18 & 0.8 & 1.6 & 0 & Y \\
SSV59   & 18 & 0.2 & 0.5 & I & Y \\
HH24MMS & 12 & 2.9 & 7.9 & 0 & Y \\
SSV63   & 12 & 1.3 & 2.3 & I & Y \\
\end{tabular}
\end{center}
\end{table}

\begin{table*}
\begin{center}
\caption[]{Best fit parameters for Gaussian (columns 2--7) and
power-law (columns 8--11) oblate models for each object modelled. Note
that the mass estimates have been corrected for helium by multiplying
the H$_2$ results by 1.25. For the Gaussian model $c_{\rm n}$
represents the maximum density and $d_{\rm n}$ and $e_{\rm n}$ the
radii (to 1/e) on the major and minor axes. For comparison, the HWHM beam radius at 450 micron is 0.008pc.The $i_{\rm z}$ and
$i_{\rm r}$ columns show the indices for the power law n$_{\rm H_2}$
distribution in each direction (i.e. when $\theta=\pi$ and $\theta=0$
respectively in equation~\ref{eq:cloudh1}).  $c_{\rm n}$ is then the
density at the inner cut-off radius 0.001pc. The temperature
distribution has been assumed to be spherically symmetric and varies
according to $T_{\rm K}=a_TR^{0.3}+b_TR^{-0.3}$ where $R$ is the
distance to the centre of the cloud measured in parsecs, $a_T$ being
zero for the power law models.
\label{tab:bestfit}}
\begin{tabular}{lcccccccccccc}
Object &  Mass     & Density & \multicolumn{2}{c}{Scale} &
       \multicolumn{2}{c}{Temperature (K)} & Density & \multicolumn{2}{c}{Index} & Temperature (K) \\
       & (M$_\odot$) & $c_n$ (cm$^{-3}$)  & $d_n$ (pc) & $e_n$ (pc) & $a_T$ & $b_T$ & $c_n$ (cm$^{-3}$) &
       $i_z$  & $i_r$ & $b_T$ \\ 
\hline
LBS17H & 16 & 2.4$\times10^7$& 0.014 & 0.010 & 10 & 1.6 & 1.1$\times10^8$ & --1.4 & --1.2 & 2.2 \\
LBS18  & 21 & 9.8$\times10^6$& 0.025 & 0.010 &  0 & 2.1 & 2.3$\times10^8$ & --1.7 & --1.3 & 1.7 \\
HH24MMS& 50 & 2.6$\times10^8$& 0.009 & 0.007 & 15 & 1.0 & 5.6$\times10^8$ & --1.6 & --1.2 & 1.9 \\
HH25MMS& 12 & 1.4$\times10^6$& 0.045 & 0.018 &  5 & 4.3 & 1.3$\times10^7$ & --1.1 & --0.7 & 3.9 \\
SSV63  &  8 & 1.9$\times10^6$& 0.030 & 0.015 & 10 & 2.9 & 3.0$\times10^7$ & --1.6 & --1.3 & 3.0 \\
\end{tabular}
\end{center}
\end{table*}

\section{Analysis}

\subsection{Simplest model -- spherically-symmetric, isothermal cores}

To determine the mass and temperature of the clumps requires a
knowledge of the emissivity law, and that the distribution of emission
should be well resolved and observed with high signal-to-noise ratio
at at least two different wavelengths. These conditions are not
satisfied for our observations. The emissivity law is uncertain and
even if this were not the case uncertainty in calibration (dominated
by the 450 $\mu$m measurements) would be a strongly limiting factor in
fixing the temperature and hence the mass. The clumps are barely
resolved and the temperature within them may well vary with position,
so that only an oddly weighted beam average can be derived. How then
are we to proceed?

Initially we have assumed the clumps possess the simplest possible
structure. They were assumed to be optically thin and at a constant
temperature, which we obtain from previous ammonia observations
(Menten, Walmsley \& Mauersberger 1987; Verdes-Montenegro \& Ho
1996). Where no prior temperature data exist we have assumed an
intermediate value of 15~K. Better estimates of the dust temperature
may be obtained from greybody fitting if enough data exist. We
employed the Hildebrand~\shortcite{hildebrand} emissivity value of
0.1~cm$^2$g$^{-1}$ at 250~$\mu$m to estimate the appropriate values at
850~$\mu$m and 450~$\mu$m assuming a value of $\beta$ of 1.5, the
median value typical of dust emission in dense cloud cores
(e.g. Huard, Sandell \& Weintraub 1999; Chandler \& Richer 2000).  The
core mass is given by $F_\nu d^2/(\kappa_\nu B_{\nu, T})$ where
$F_\nu$ is the flux density at frequency $\nu$, $d$ is the distance to
the source, $\kappa_\nu$ is the absorption coefficient and $B_{\nu,
T}$ is the Planck function for a blackbody of temperature $T$.

Even if these assumptions closely model reality, the calibration error
alone limits the determination of mass to within 50 to 100 per
cent. The assumed temperatures and derived masses are shown in
Table~\ref{tab:simpleanalysis}. Despite the uncertainties, it seems
that the dust cores probably have a mass within a 0.03 pc diameter
region of order 1--2~M$_\odot$. The mass derived for HH24MMS seems
anomalously high, which led Chini et al. (1993) to suggest that the
absorption coefficient is larger towards this source (possibly due to
a lower value for $\beta$ or a coagulation of dust grains). 

%Note that had we assumed a larger value of $\beta$ the derived masses
%would have been larger due to the smaller predicted absorption
%coefficient at submillimetre wavelengths.

%To derive the temperature the 450 $\mu$m map was convolved to the same
%resolution as that at 850 $\mu$m (15 arcsec). The temperature was
%determined from the flux ratio and hence the mass within the 15 arcsec
%beam.

%For temperatures above about 20 K, the flux density ratio is
%insensitive to temperature and becomes instead determined by the value
%of $\beta$. For three clumps it is possible to make a comparison
%between gas kinetic temperatures derived from ammonia and dust
%temperatures from continuum emission. In SSV63 $T({\rm
%NH}_3)\sim$11--15 K, $T_{\rm dust} \sim$ 9--20 K, in HH24MMS $T({\rm
%NH}_3)\sim$11--15 K, $T_{\rm dust}\sim$7--10 K, and in HH25MMS $T({\rm
%NH}_3)\sim$17 K, $T_{\rm dust}\sim$10--23 K. The agreement for the
%temperature between the ammonia values and those derived from the dust
%is reasonably good. It suggests that the 30 percent estimate of
%systematic calibration error for the 450 $\mu$m fluxes used in the
%calculations may be somewhat conservative, and perhaps that the
%assumption of a $\beta$ value of 2 (implicit in the Hildebrand
%formulation) is not too outlandish.

We have made greybody fits to the spectral energy distributions using
our SCUBA and IRAS HiRes-processed data to place an upper limit on the
dust temperature (see Gibb \& Little 2000, Gibb \& Davis 1998 for
greybody fits to two of our SCUBA sources). In general this
temperature lies between 20 and 30 K, depending on the value of
$\beta$ for the fit, with the clumps containing IRAS point sources
tending to have higher upper limits than those dust clumps with no
IRAS point source. Unfortunately the greybody fits cannot provide more
accurate estimates of the temperature because the IRAS fluxes are
measured with a beam considerably larger than that of our JCMT
observations, one which often encompasses more than one of our SCUBA
clumps.

If a power law
variation of density with radius were assumed and the temperature
constant this would imply a power-law index of --1.7, a value
intermediate between the value expected for a singular isothermal
sphere in hydrostatic equilibrium (--2) and one in free-fall collapse
(--1.5). If temperature increases inward the power law would be
flatter. If a Gaussian distribution is assumed, an angular diameter
for the sources $\sim$12 arcsec (0.023 pc) is indicated, probably too
close to the SCUBA beamsize to be meaningful.

\subsection{More detailed modelling of the emission}

Although the data indicate that the observations have only partially
resolved most of the sources (particularly in the narrow direction)
the existence of simultaneously observed well registered maps at both
850 and 450 $\mu$ms justified more detailed modelling which would be
capable of taking into account density and temperature gradients and
finite optical depth. In particular we wished to compare the
predictions of both Gaussian and power-law models, as they represent
popular paradigms for the structure of such objects. We might attempt
to distinguish between power law and Gaussian distributions by fitting
the 850 $\mu$m and 450 $\mu$m observations {\em simultaneously}.

Accordingly, we have written a program that computes fluxes for model
clouds with axial symmetry, described in the Appendix.  To compare the
model output with the telescope data two cuts were made on each of the
SCUBA maps, one down the long axis of the cloud and one perpendicular
to this. These cuts were then repeated on the model output and the
results compared. The best fit oblate Gaussian and power law 
models for each object modelled are presented in
%Tables~\ref{tab:bestfit1}~\&~\ref{tab:bestfit2}. 
Table~\ref{tab:bestfit}.
Given the uncertainties in the calibration, which will directly propagate
into the magnitude of the density and temperature in the model, we
emphasize that in this section we seek to explore the possible forms of
the density variation, rather than absolute values. 

We tried both the Hildebrand ($\beta$=2) and Testi and Sargent (1998) 
($\beta$=1.1) versions of the emissivity function (see Appendix) to see
whether the choice of $\beta$ affects our results significantly. For
the Gaussian distribution there is no significant observable change in
the any of the output cross-sections. However, it is necessary to
decrease the density by a factor of approximately 2 and increase
temperature by a factor of approximately 1.5 in order to retain the
correct intensities which has the result of decreasing the mass of the
cloud by around 50 per cent. Therefore there does not seem to be any
clear way of distinguishing between different values of $\beta$ using
this data set.

We have also investigated the differences caused by assuming an oblate or
a prolate spheroid as the cloud shape. Unfortunately there does not seem
to be any way of distinguishing between the two models as the best fit
line profiles are virtually identical in both cases. However, the
parameters needed to obtain the best fit are significantly different. The
Gaussian prolate case needs a higher density by a factor of between 1.3
and 3 (the more elliptical the cloud, the larger the difference). This in
turn causes a higher cloud mass (up to 15 per cent). There is also a
corresponding slight decrease in the temperatures required. All other
parameters remain the same. For the power law distributions the same
comments apply except that the increase in density caused a larger
increase in the cloud mass (up to 50 per cent). 

\subsubsection{Features of the modelling}

The results of the model fits are given in
%Tables~\ref{tab:bestfit1}~\&~\ref{tab:bestfit2}. 
Table~\ref{tab:bestfit}.
In attempting the fits it was found that:

        (1) At least two of the sources (LBS18S and HH25MMS) showed 
evidence for two or more components along their major axes.

        (2) There was usually evidence of extended `ridge' or `wing'
emission (up to 25 per cent of the peak) which could not be accounted
for in the modelling carried out here.  

        (3) Gaussian fits are usually slightly better than power laws in
one significant respect, described below.

\begin{figure*}
\includegraphics*[scale=0.75]{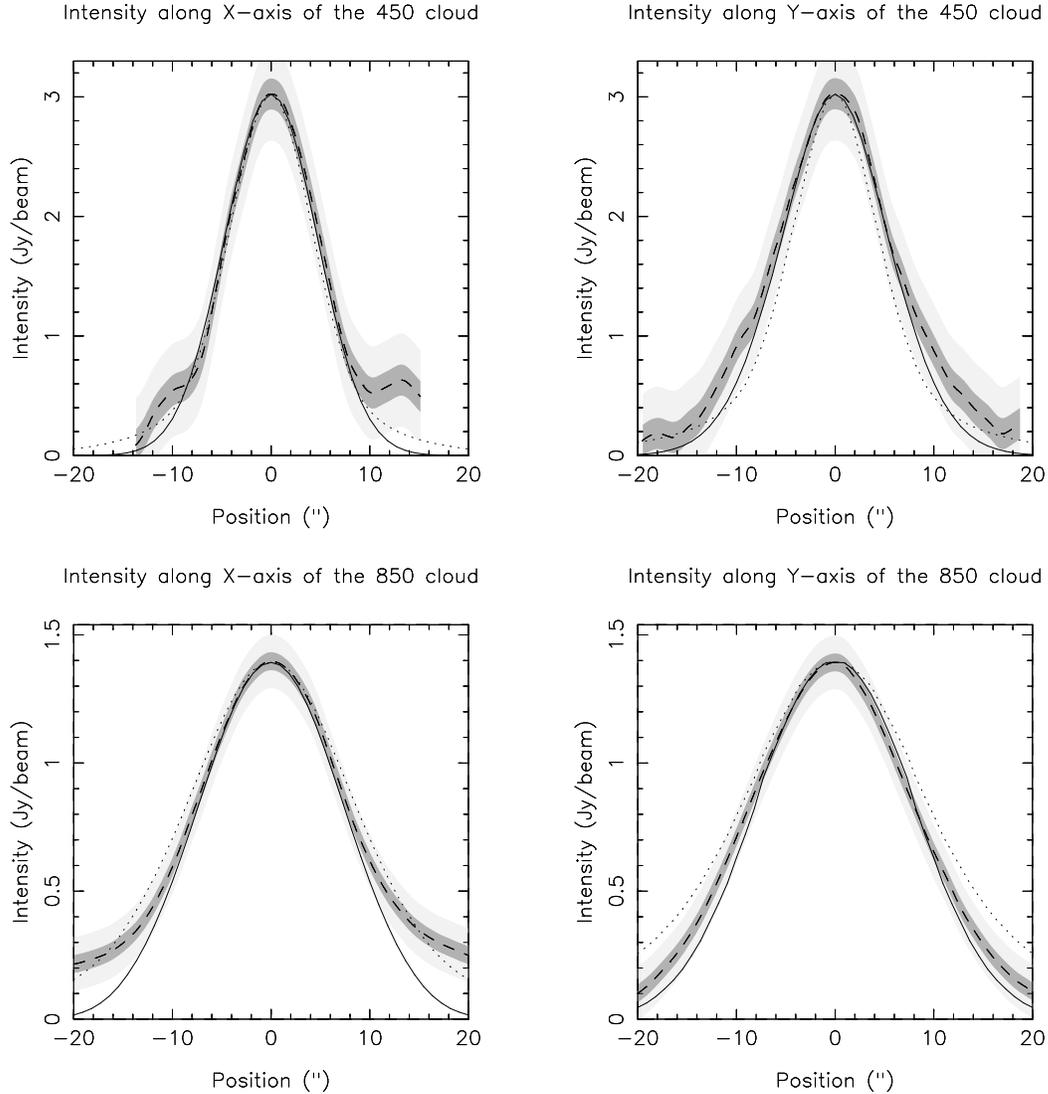}
\caption[]{Cuts along the major and minor axes of LBS17H. The solid
and dotted lines are respectively the modelled results from a Gaussian
density distribution and a power law distribution.  The dashed lines
are cuts taken from Fig.~\ref{fig:lbs17}. The %dark and light grey
strips are 
the 1$\sigma$ and 
3$\sigma$ noise levels 
(from Fig.~\ref{fig:lbs17}).  
The beamsizes used were 8 arcsec for 450 $\mu$m and 15 arcsec for 850
$\mu$m. \label{fig:lbs17model}}
\end{figure*}

Fig.~\ref{fig:lbs17model} shows the fits for LBS17. The presence of a
ridge connecting the main clump to a secondary clump to the north west
causes an extended low level emission wing in the 850 $\mu$m $x$-axis
cut which no effort has been made to fit. Ignoring this, the central
part of the $x$-axis cut and the whole of the $y$-axis cut is almost
perfectly fit by a Gaussian distribution profile. By contrast we were
unable to find a power law fit that successfully modelled both the
450 $\mu$m and 850 $\mu$m maps {\em simultaneously}.

For power laws the 850 $\mu$m $y$-axis cut
cannot be made narrow enough without causing the 450 $\mu$m $y$-axis
cut to become too narrow. The smaller 450 $\mu$m beam weights the
central material more heavily -- too heavily for the inwardly
increasing power law. 

Of the five sources which were modelled this feature was observed
along the major axes of SSV63, LBS17H, and LBS18S, and the minor axes
of HH25MMS, HH24MMS, and LBS17H. Although the effect is not large it
is noticeable and we believe it to be a significant discriminant. This
would suggest that the density law flattens towards the centre of the
clump (as in a Gaussian), although thus far such a requirement has
only been determined for pre-collapse objects (e.g. Ward-Thompson et
al. 1994). Future higher-resolution observations are clearly necessary
to determine the density variation in the inner regions of these
clumps.

Investigation on dummy data showed that the effect of fitting a single
component model to a two component source (i.e. main component plus
`ridge') was to produce a behaviour in the opposite sense to that
observed. The one object where a power law seems to fit reasonably
well (SSV63) has a less pronounced peak on the map
(Fig.~\ref{fig:lbs23n}) than the others.

\section{Discussion}

\subsection{Dust temperatures}

From simple isothermal greybody modelling the derived dust
temperatures in LBS23 are generally consistent with those derived from
HCO$^+$ and NH$_3$ observations except in the case of HH24MMS, which
yields a low dust temperature (Table~\ref{tab:simpleanalysis}). As
mentioned above, Chini et al. (1983) attribute this to an enhanced
absorption coefficient. However, it should be noted that greybody
fitting of the spectral energy distribution yields more accurate
results as with enough data, the emissivity and its frequency
dependence may be constrained in addition to the temperature
(e.g. Ward-Thompson et al. 1995).

The dust temperatures tend to be rather lower than the CO-derived
temperatures, which, on account of the very high CO optical depth, are
likely to refer to the outer parts of the cloud. Although external
heating of these outer parts therefore seems likely, it should be
noted that the temperatures estimated for the ridges of LBS17 and
LBS18 show little or no evidence of it. This fact indicates that the
majority of the dust emission arises from the cold interior of the
cores observed here.
  
\subsection{Comparison with protostellar models}

The canonical protostellar collapse model of Shu~\shortcite{shu}
predicts inside-out collapse in an isothermal spherical envelope,
leading to a density distribution with a power law of index --1.5 in
the collapse region while outside it the index has the isothermal
value of --2.0. The effect of magnetic fields and/or rotation will
obviously complicate this simple picture. For example, the simulations
of Crutcher~et~al.~\shortcite{cmt}, which include the effects of
magnetic fields and ambipolar diffusion, indicate modifications of
these laws to --1.3 and --1.7. Flattening would be expected leading to
elongation in projection onto the plane of the sky.

The observations of the clumps do indeed reveal elongation which is
particularly evident at 450 $\mu$m where the 8 arcsec beamwidth of the
telescope is equivalent to 3200 AU. However, HH25MMS, HH24MMS and
LBS18 are elongated in the direction of the filamentary structure in
which they are embedded, suggesting perhaps that they might be
prolate. In SSV63 and LBS17H the elongation appears to be
perpendicular to the filaments.

When the observable extent of a source is not much greater than the
beamwidth it is perhaps invidious to seek to distinguish between
Gaussians and power laws. The central regions of the clumps can be
fitted by Gaussians. It is hard to separate wing emission from a
clump from emission due to the embedding filament. It is clear that,
if the densities are interpreted as power law distributions, the power
law indices vary with direction. The indices lie in the range --0.7
to --1.2 in the direction of elongation and --1.1 to --1.7
perpendicular to it. These values have been derived assuming an
$r^{-0.3}$ temperature dependence. If the clumps were taken to be
isothermal their magnitude would increase by about 0.3. Given that in
the central regions Gaussians are somewhat favoured anyway, the least
that can be said is that the results do not agree very well with the
simple Shu model.

Questions pertinent to our results include: (a) are the cores oblate or
prolate?, and (b) what has led to their orientation relative to the filaments
in which they are embedded?  While a number of papers have been published
examining the fragmentation of filaments (e.g. Fiege \& Pudritz 2000;
Nakajima \& Hanawa 1996), few have attempted to analyze the structure of
the fragments and in particular look at the issue of prolate or oblate
cores. However, Bonnell, Bate~\&~Price (1996; hereafter BBP) have
performed computer simulations of the dynamic evolution of initially
prolate, magnetic field- and rotation-free filaments into protostars,
which may be usefully related to our observations. The evolution depends
primarily on the ratios of the Jeans radius to the minor and major axes of
the filament, and they show that under certain conditions it is possible
to produce an oblate core embedded within a prolate condensation.

BBP give arguments to suggest that rotation would be unlikely to modify
hugely these results. However, despite the claims of BBP, it is likely
that magnetic fields may do so although their effects would depend on the
field strength, direction and the Alfv\'en speed. For example, if the
field lines were perpendicular to the axis of the filament then collapse
across the field lines towards a single object might be inhibited on
dynamical timescales, resulting in a string of individual cores which each 
form a protostar.  

Can these theoretical considerations be related to our observations?
First, we note that the effect of projecting both prolate and oblate
object onto the plane of the sky is to decrease their apparent asymmetry. 
Both the filaments and the envelopes are, on average, more asymmetric than
they appear. Second, the typical core size of a few $\times$10$^3$ au is
similar to the scale of the flattened structures in BBP's simulations.
Thirdly, like LBS23 and NGC2024, LBS17 and LBS18 both have an elongated
filamentary structure with protostellar cores appearing to form as a
result of fragmentation along their length. Within the framework of BBP's
simulations, their original filamentary structure was sufficiently massive
that fragmentation has occurred along their length, as may have also
happened in LBS18 (see Fig. 4). One would therefore expect that the
condensations in these objects should be regarded as BBP's `intermediate'
prolate objects aligned with the original filaments. 

HH25MMS, HH24MMS and LBS18S are elongated in the direction of the
filamentary structure in which they are embedded, suggesting perhaps
that they might indeed be prolate `intermediate' cores which form
after fragmentation of a prolate filament (BBP). But in SSV63 and
LBS17H the elongation appears to be perpendicular to the filaments --
within the framework of BBP's analysis they would be oblate `ultimate'
cores. Both these cores are observed at the ends, rather than
centres, of filaments, however. Could it be that they have `swallowed'
all the material on one side of them? However, LBS18S is also observed
at the end of a filament, but is elongated along the filament.

As there is evidence of star formation all along LBS23 it seems likely
that protostars have sufficient time to form within the intermediate
envelopes.  Their mutual attraction could lead finally to a stellar
cluster -- not BBP's merging into a single object which results from the
symmetry of their initial starting conditions. BBP also note that the
initial conditions require the accumulation of many Jeans masses in a time
much shorter than the dynamic time scale, which might be achieved, for
example, by cloud-cloud collisions, as proposed for the formation of the
filament TMC1 by Little~et~al.~\shortcite{lrm}. 

\section{Conclusions: searching for protostars}

Searching LBS cores with SCUBA is a powerful method for detecting
protostellar candidates in the Orion clouds. Several new ones have
been discovered in very little observing time. On the simplest
interpretation the clumps observed have masses (in the 15 arcsec beam)
varying from a few (SSV59, SSV63, HH26) to 50 (HH24MMS) solar
masses. Of the newly detected sources SSV59 and HH26IR are probably
Class I objects, while LBS17H is a new Class 0 source. LBS18S has no
near-infrared, IRAS or radio source, nor does it appear to have an
outflow and so it is very young, perhaps pre-protostellar. Further
studies of these objects, e.g. by interferometry, are required to
confirm their evolutionary status.

One (LBS18) out of the 4 small LBS cores shows evidence of a
protostellar clump of several solar masses.  There are two in LBS17
(including the 1300 $\mu$m source of Launhardt et al. 1996) and
several in LBS23. On the basis of present limited observations (4
small LBS cores, 2 large LBS cores) the ratio of `protostellar' clump
mass to the total virial mass of the filament does seem not very
different for large or small cores -- but the statistics are not
good. A proper census of the LBS cores, in which both small and large
cores are mapped in their entirety, feasible with SCUBA, will allow a
mass spectrum of these clumps to be determined. The next obvious step
is to carry out a wide field, unbiased survey of L\,1630 to determine
the mass distribution on 15-arcsec scales.

\vspace{5mm}
\noindent
{\bf ACKNOWLEDGMENTS}\\

\noindent
The authors would like to thank the service observers for their
efforts in obtaining these data, and Wayne Holland for his help in
resolving early problems with the data.

\label{lastpage}

\appendix

\section{Dust radiative transfer model}

Our radiative transfer model is a simple modification to the
axisymmetric molecular line transfer code of Phillips \& Little
(2000) to cope with dust continuum emission.

The emissivity, $E_\nu$, of a small element $i$ is written as
\begin{equation}
E_{\nu,i}=\kappa_{\nu,i} \frac{2 h \nu^3}{c^2(e^{\frac{h\nu}{kT_{\rm d}}}-1)}
\end{equation}
where the dust is at a temperature $T_{\rm d}$.  The cloud to be
modelled is split into geometrical subsections. The geometrical
subsections chosen for the cloud are a series of stacked concentric
rings. Each subsection is assigned a temperature and a molecular
hydrogen density according to the desired model. For any desired line
of sight the positions of the intersections of the line of sight with
the boundaries of the subsections are then calculated. It is then
possible to integrate the emission along the line of sight simply by
adding up the contributions from each segment on the line of sight
($\Delta l_i$) according to
\begin{equation}
I_\nu=\sum_{i=1}^{n}E_{\nu,i} \Delta l_i e^{-\sum_{j=1}^{i}\tau_{\nu,j}}
\end{equation}
where $E_{\nu,i}$ and $\tau_{\nu,j}$ are the emissivities and optical
depths for segment $i$ (or $j$) respectively.

The beam-averaged emission is computed by taking a grid of positions
over the beam, calculating the emission from each and summing the
results with appropriate weightings. For all the modelling that
follows the beam is assumed to be a perfect Gaussian. Whilst this is a
good assumption for the 850 $\mu$m data the JCMT beam deviates from a
perfect Gaussian at 450 $\mu$m due to imperfections in the
dish. Observations of Uranus showed that the beam had a substantial
`error' component containing $\sim$50 per cent of the power and spread
over a 50 arcsec-diameter region. However, although the precise shapes
of the 450 $\mu$m cuts may be slightly inaccurate (mostly at the lower
levels), the difference is almost certainly not enough to
significantly affect our results.

Two different models for the molecular hydrogen distribution were
tested. The first is a power law model of the form
\begin{equation} \label{eq:cloudh1}
%n_{\rm H_2}=c_n\left( \frac{R}{R_{\rm in}}\right) ^{-d_n-e_n \left[ \cos
n_{\rm H_2}=c_n D^{-d_n-e_n \left[ \cos %original version above
\left( \frac{\pi}{2}\left( \frac{2 \left| \theta
\right|}{\pi}\right)^{m}\right) \right]^2}
\end{equation}
where 
$D$=$R/R_{\rm in}$, %get rid of this if you like...
$\theta=\tan^{-1} \left( \frac{z}{r} \right)$ and $r$ and $z$
are the cylindrical co-ordinates of the position of interest in the
cloud with the $z$-axis being the axis of symmetry in the cloud and
$r$ the distance from the axis, while $R$ refers to the distance from
the cloud centre measured in parsecs. The $z$ axis is taken to lie in
the plane of the sky. Choosing appropriate values of $d_n$ and $e_n$
enables the power law dependence to vary with direction, thus allowing
an elongated cloud to be described. By changing the sign of $e$ it is
possible to model either a prolate or an oblate cloud.  Although such
a description has the major advantage of enabling easy comparisons
with various predictions for the power law in each direction it is
difficult to obtain smooth elliptical contours of equal density. It is
necessary to give the value of $c_n$ for a position, $R_{\rm in}$,
close to the centre of the cloud rather than at the outer edge as many
authors have done (e.g.  Little~et~al.~1994) since specifying the
density at the outer edge causes the two different power laws to give
two different density values at the cloud centre. The apparently odd
angular variation simply produces nicely oval contours, with $m$
chosen between 1.3 and 1.5 for prolate clouds and between 0.7 and 0.8
for oblate ones. We also tried using a Gaussian distribution of the
form
\begin{equation} \label{eq:cloudh2} n_{\rm
H_2}=c_n \exp \left(-\left[ \left( \frac{r}{d_n}\right)^2+\left(
\frac{z}{e_n}\right)^2\right]\right)
\end{equation}
where $r$ and $z$ are as described above for the power law. The
parameters $d_n$ and $e_n$ enable the size of the Gaussian (and hence
the cloud) to be specified.

The temperature distribution has been assumed to be spherically
symmetric and varies according to
\begin{equation} \label{eq:cloudtemp}
T_{\rm K}=a_TR^{0.3}+b_TR^{-0.3}
\end{equation}
where $R$ is the distance to the centre of the cloud measured in
parsecs. This distribution, which crudely represents a cloud that may
be both internally and externally heated, is taken from the results of
Scoville~\&~Kwan~\shortcite{scokan} which assumes that $\beta =2$. GL98
applied this formula to HH24MMS and HH25MMS in LBS23 for modelling
C\,{\sc i} and C$^{18}$O. In principle the temperature variation with
distance from the centre should be different along the two axes of the
flattened model core. However, Scoville \& Kwan~\shortcite{scokan}
show that the optical depth probably has a greater effect on the
temperature fall off than the density law, and thus for simplicity we
retain a spherically-symmetric temperature distribution.

For all models a cloud consisting of 60 cylinders and 60 disks (i.e. a
total of 3540 rings) was used with both the disks and cylinders packed
closer together near the centre of the cloud (this enables the rapid
changes of the power laws near the centre of the cloud to be modelled
without large jumps in the parameter values from one ring to
another). The cylinders have an outer radius of 0.05 pc and the disks
extend out to a distance of 0.05 pc above and below the centre of the
cloud. The exact choice of this distance does not affect the results
for the Gaussian models. However, power law models are affected to
varying degrees depending upon the steepness of the power law
used. For steep power laws the cross-section shape remains
approximately the same with only a 10 per cent change in $c_n$ when
the radius is increased by a factor of 4. For the shallow power law
the changes are much bigger. For the worst case increasing the radius
of the cloud by a factor of 4 leads to significantly higher emission
at the edge of the cross sections, particularly for the 450 $\mu$m
crossection. This can be corrected by making the power law a little
steeper (0.1 extra in each direction) with a 60 per cent increase in
$c_n$.

In order to model the beam correctly for the rapid changes in the
centre of the cloud a total of up to 55$\times$55=3025 lines of
sight are distributed across the beam. This number was determined by
trying progressively larger numbers of lines of sight for each model
until the cross-sections stopped changing.

The program initially calculates the intensity only for the central
point in the cloud and then, by adjusting the values of $b_T$ (in
equation~\ref{eq:cloudtemp}) and $c_n$ (in
equations~\ref{eq:cloudh1}~\&~\ref{eq:cloudh2}), it fits the peak
temperatures in the 450 and 850 $\mu$m maps. Once the peaks are fitted
correctly a map is produced and then the four cuts are compared with
the telescope data.

With power laws there is usually a problem with deciding where to
place an inner cut off $R_{\rm in}$ (necessary to prevent a singularity
at the centre of the cloud). We have used 0.001 pc here,
although there are only minor changes in the results if this is
increased or decreased by a factor of 5. This is because the extremely
small area in the centre of the cloud has little affect on the total
emission over the beam.

There is also some question as to the precise form of the emissivity.
The Hildebrand~\shortcite{hildebrand} formula, which we have used
here, proposes that in the long wavelength regime ($\lambda > 250
\mu$m) the absorption coefficient at a frequency $\nu$ (GHz) is
related to the molecular hydrogen density $n_{{\rm H}_2}$ (cm$^{-3}$)
by
\begin{equation} \label{eq:hildebrand}
\kappa_\nu=\frac{2 n_{{\rm H}_2}} {1.2 \times 10^{25}}\left(
\frac{\nu}{750}\right)^2 
\hspace{1cm} {\rm cm^{-1}} 
\end{equation}
Others (e.g. Visser~et~al.~1998 and Testi~ \&~Sargent~1998) have
recently suggested that lower values for $\beta$ may be more appropriate. We
tested this by using the relationship suggested by
Testi~\&~Sargent~\shortcite{testsarg} of $\kappa_\nu= \frac{2 n_{{\rm
H_2}}}{9.6 \times 10^{25}}\left( \frac{\nu}{230}\right) ^{1.1}$
cm$^{-1}$ in place of equation~\ref{eq:hildebrand}, where as before
$\nu$ is in GHz and $n_{\rm H_2}$ is in cm$^{-3}$. However, as
mentioned in the text, the precise choice of emissivity law does not
significantly affect the results (other than the magnitude of the
temperature and density).

\bsp

\begin{thebibliography}{}
\bibitem[\protect\citename{Andr\'e, Ward-Thompson \& Barsony }1993]{awb93}
Andr\'e P., Ward-Thompson D., Barsony M., 1993, ApJ, 406, 122
\bibitem[\protect\citename{Bonnell, Bate \& Price }1996]{bbp}
Bonnell I.A., Bate M.R., Price N.M., 1996,
MNRAS, 279, 121 (BBP)
\bibitem[\protect\citename{Bontemps, Ward-Thompson \& Andr\'e }1996]{bwa}
Bontemps S., Ward-Thompson D., Andr\'e P., 1996,
A\&A, 314, 477
%\bibitem[\protect\citename{Boss \& Myhill }1995]{boss}
%Boss A.P., Myhill E.A., 1995,
%ApJ, 451, 218
\bibitem[\protect\citename{Chandler \& Carlstrom }1996]{cjc}
Chandler C.J., Carlstrom J.E., 1996,
ApJ, 466, 338
\bibitem[\protect\citename{Chandler, Moore \& Emerson }1992]{cme}
Chandler C.J., Moore T.J.T., Emerson J.P., 1992,
MNRAS, 256, 369
\bibitem[\protect\citename{Chandler \& Richer }2000]{cr00}
Chandler C.J., Richer J.S., 2000,
ApJ, 530, 851
\bibitem[\protect\citename{Chini et al. }1993]{ckh}
Chini R., Krugel E., Haslam, C.G.T., Kreysa E., Lemke R., Reipurth B., 
Sievers A., Ward-Thompson D., 1993,
A\&A, 272, L5
\bibitem[\protect\citename{Choi, Panis \& Evans }1999]{choi}
Choi M., Panis J-F., Evans N.J., II, 1999,
ApJS, in press
\bibitem[\protect\citename{Crutcher et al. }1994]{cmt}
Crutcher R.M., Mouschovias T.Ch., Troland T.H., Ciolek G.E., 1994,
ApJ, 427,839
\bibitem[\protect\citename{Davis et al. }1997]{d97}
Davis C.J., Ray T.P., Eisl\"offel J., Corcoran D., 1997,
A\&A, 324, 263
\bibitem[\protect\citename{Dent, Matthews \& Walther }1995]{dent}
Dent W.R.F., Matthews H.E., Walther D.M., 1995,
MNRAS, 277, 193
\bibitem[\protect\citename{}]{fp}
Fiege J.D., Pudritz R.E., 2000,
MNRAS, 311, 105
\bibitem[\protect\citename{Gibb }1999]{g99}
Gibb A.G., 1999,
MNRAS, 304, 1
\bibitem[\protect\citename{Gibb \& Davis }1997]{gd97}
Gibb A.G., Davis C.J., 1997,
in Malbet F., Castets A., eds, Poster Summaries, IAU Symp. 182, Low Mass Star Formation -- from Infall
to Outflow. Universit\'e Joseph Fourier, Grenoble, p. 120
\bibitem[\protect\citename{Gibb \& Davis }1998]{gd98}
Gibb A.G., Davis C.J., 1998,
MNRAS, 298, 644
\bibitem[\protect\citename{Gibb \& Heaton }1993]{gh}
Gibb A.G., Heaton B.D., 1993,
A\&A, 276, 511 (GH93)
\bibitem[\protect\citename{Gibb \& Little }1998]{gl98}
Gibb A.G., Little L.T., 1998,
MNRAS, 295, 299 (GL98)
\bibitem[\protect\citename{Gibb \& Little }2000]{gl2}
Gibb A.G., Little L.T., 2000,
MNRAS, 313, 663
\bibitem[\protect\citename{Gibb et al. }1995]{glhl}
Gibb A.G., Little L.T., Heaton B.D., Lehtinen K.K., 1995,
MNRAS, 277, 341  (Paper I)
\bibitem[\protect\citename{Haschick, Moran \& Rodr\'{\i}guez }]{haschick} 
Haschick A.D., Moran J.M., Rodr\'{\i}guez L.F., Ho P.T.P.,
1983, ApJ, 265, 281
\bibitem[\protect\citename{Hildebrand }1983]{hildebrand}
Hildebrand R.H., 1983,
QJRAS, 152, 1
\bibitem[\protect\citename{Holland et al. }1999]{crg}
Holland W.S., Robson E.I., Gear W.K., Cunningham C.R., Lightfoot J.F.,
Jenness T., Ivison R.J., Stevens J.A., Ade P.A.R., Griffin M.J., 
Murphy J.A., Naylor D.A., 1999, 
MNRAS, 303, 659 
\bibitem[\protect\citename{Huard et al., }1999]{hsw}
Huard T.L., Sandell G., Weintraub D.A., 1999, ApJ, 526, 833
\bibitem[\protect\citename{The Physics of Star Formation }1991]{l91}
Lada C.J., 1991, in Lada C.J., Kylafis N.D., eds, The Physics of Star
Formation and Early Stellar Evolution, Kluwer, Dordrecht, p. 329
\bibitem[\protect\citename{Lada, Bally \& Stark}1991]{lbs}
Lada E.A., Bally J., Stark A.A., 1991,
ApJ, 368, 432
\bibitem[\protect\citename{Launhardt et al. }1996]{lmh}
Launhardt R., Mezger P.G., Haslam C.G.T., Kreysa E., Lemke R., 
Sievers A., Zylka R., 1996, 
A\&A, 312, 569
\bibitem[\protect\citename{Lis, Carlstrom \& Phillips }1991]{lcp}
Lis D.C., Carlstrom J.E., Phillips T.G., 1991,
ApJ, 370, 583
\bibitem[\protect\citename{Lis, Menten \& Zylka }1999]{lmz}
Lis D.C., Menten K.M., Zylka R., 1999,
ApJ, 527, 856
\bibitem[\protect\citename{Little et al. }1978]{lrm}
Little L.T., Riley P.W. Macdonald G.H., Matheson D.N, 1978,
MNRAS, 183, 805
\bibitem[\protect\citename{Little et al. }1994]{lg1}
Little L.T., Gibb A.G., Heaton B.D., Ellison B.N., Claude S.M.X., 1994,
MNRAS, 217, 649
\bibitem[\protect\citename{Mauersberger et al. }1992]{mwm}
Mauersberger R., Wilson T.L., Mezger P.G., Gaume R., Johnston K.J., 1992,
A\&A, 256, 640
\bibitem[\protect\citename{Menten, Walmsley \&
Mauersberger}1987]{mwm87}
Menten K.M., Walmsley C.M., Mauersberger R., 1987, in Appenzeller I.,
Jordan C., eds, Circumstellar Matter, Reidel, Dordrecht, p.179
\bibitem[\protect\citename{Mezger et al. }1992]{msh}
Mezger P.G., Sievers A.W., Haslam C.G.T., Kreysa E., Lemke R., 
Mauersberger R., Wilson T.L., 1992, 
A\&A, 256, 631
\bibitem[\protect\citename{}]{nh}
Nakajima Y., Hanawa T., 1996,
ApJ, 467, 321
\bibitem[\protect\citename{Phillips \& Little}2000]{pl}
Phillips R.R., Little L.T., 2000,
MNRAS, 317, 179
\bibitem[\protect\citename{Scoville \& Kwan}1976]{scokan}
Scoville N.Z., Kwan J., 1976,
ApJ, 214, 488
\bibitem[\protect\citename{Shu }1977]{shu}
Shu F.H., 1977,
ApJ, 214, 488
\bibitem[\protect\citename{Testi \& Sargent }1998]{testsarg}
Testi L., Sargent A.I., 1998,
ApJ, 508, L94
\bibitem[\protect\citename{Verdes-Montenegro \& Ho }1996]{vmho}
Verdes-Montenegro L., Ho P.T.P., 1996, ApJ, 473, 929
\bibitem[\protect\citename{Visser et al. }1998]{visser}
Visser A.E., Richer J.S., Chandler C.J., Padman R., 1998,
MNRAS, 301, 585
\bibitem[\protect\citename{}1994]{wsha}
Ward-Thompson D., Scott P.F., Hills R.E., Andr\'e P., 1994,
MNRAS, 268, 276
\bibitem[\protect\citename{}1995]{wck}
Ward-Thompson D., Chini R., Kr\"ugel E., Andr\'e P., Bontemps S., 1995,
MNRAS, 274, 1219
\bibitem[\protect\citename{Zhou et al. }1991]{zhou}
Zhou S., Evans N.J., II, G\"usten R. Mundy L.G., Kutner M.L., 1991,
ApJ, 372, 518
\end{thebibliography}
\end{document}